\pdfoutput=1 
\documentclass[onecolumn]{article}

\usepackage[utf8]{inputenc}
\usepackage{graphicx}
\usepackage{authblk}
\expandafter\let\csname equation*\endcsname=\relax
\expandafter\let\csname endequation*\endcsname=\relax
\usepackage{amsmath}
\usepackage{bm}
\usepackage{xcolor}
\usepackage{braket}
\usepackage{caption}
\usepackage{subcaption}
\usepackage{geometry}
\usepackage{float}
\usepackage{array}
\usepackage{changes}
\usepackage{amssymb}
\usepackage{comment}
\usepackage{arydshln}

\title{Atom interferometry in an Einstein Elevator}

\author[1]{C. Pelluet}
\author[1]{R. Arguel}
\author[1]{M. Rabault}
\author[1]{V. Jarlaud}
\author[1]{C. Métayer}
\author[2]{B. Barrett}
\author[1,3,4,5]{P. Bouyer}
\author[1]{B. Battelier}

\affil[1]{LP2N, Laboratoire Photonique Numérique et Nanosciences, Université de Bordeaux, IOGS and CNRS , 1 Rue François Mitterrand, 33400 Talence, France}
\affil[2]{Dept.~of Physics, University of New Brunswick, 8 Bailey Dr., Fredericton NB, E3B 5A3, Canada}
\affil[3]{Van der Waals-Zeeman Institute, Institute of Physics, University of Amsterdam, Science Park 904, 1098 XH,
Amsterdam, The Netherlands}
\affil[4]{QuSoft, Science Park 123, 1098 XG, Amsterdam, The Netherlands}
\affil[5]{Eindhoven University of Technology, P.O.~Box 513, 5600 MB, Eindhoven, The Netherlands}

\begin{document}
\maketitle

\begin{abstract}
Recent advances in atom interferometry have led to the development of quantum inertial sensors with outstanding performance in terms of sensitivity, accuracy, and long-term stability. For ground-based implementations, these sensors are ultimately limited by the free-fall height of atomic fountains required to interrogate the atoms over extended timescales. This limitation can be overcome in Space and in unique ``microgravity'' facilities such as drop towers or free-falling aircraft. These facilities require large investments, long development times, and place stringent constraints on instruments that further limit their widespread use. The available ``up time'' for experiments is also quite low, making extended studies challenging. In this work, we present a new approach in which atom interferometry is performed in a laboratory-scale Einstein Elevator. Our experiment is mounted to a moving platform that mimics the vertical free-fall trajectory every 13.5 seconds. With a total interrogation time of $2T = 200$ ms, we demonstrate an acceleration sensitivity of $6 \times 10^{-7}$ m/s$^{2}$ per shot, limited primarily by the temperature of our atomic samples. We further demonstrate the capability to perform long-term statistical studies by operating the Einstein Elevator over several days with high reproducibility. These represent state-of-the-art results achieved in microgravity and further demonstrates the potential of quantum inertial sensors in Space. Our microgravity platform is both an alternative to large atomic fountains and a versatile facility to prepare future Space missions.
\end{abstract}

\begin{figure}[!ht]
    \centering
    \includegraphics[width=0.85\textwidth]{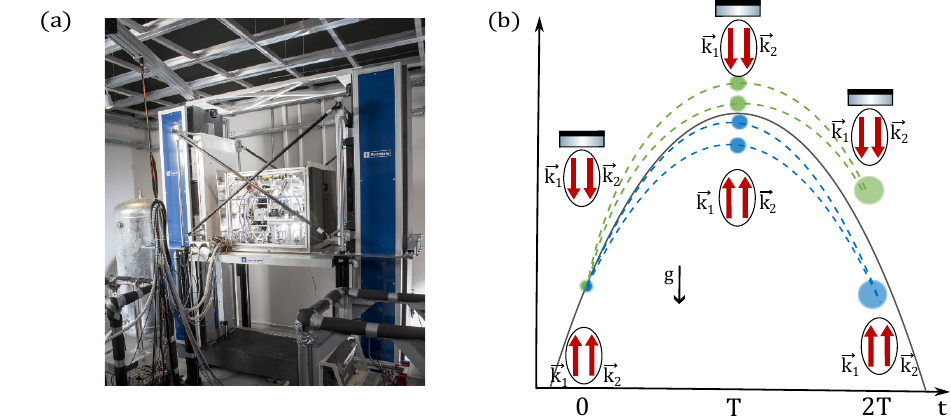}
    \caption{\textbf{Atom interferometer in weightlessness.} (a) Picture of the apparatus installed on the Einstein Elevator. (b) The experiment, including the atomic cloud, the reference mirror and the retroreflected laser beam follows the parabolic trajectory of the platform. A mechanical accelerometer (Colibrys) is attached to the reference mirror. The first Raman pulse achieves a symmetric double velocity selection and only a fraction of the atomic sample participates in the upper (green) or the lower (blue) atom interferometer. We measure simultaneously the outputs of the double interferometer ($N_2(\pm),N_{\rm T}(\pm)$) with a single photodetector leading to a single population ratio $R$ (see Eq.~\eqref{eq:OutputAIDSD}). }
    \label{fig:AtomInterferometryMicrogravity}
\end{figure}

The exceptional performance offered by quantum sensors based on light-pulse atom interferometry is based on two essential characteristics: they are well controlled and understood, and they are isolated from many environment effects \cite{Bongs2019, Geiger2020}. In the case of inertial sensors, free-falling atoms provide an ideal quantum reference. Today, different implementations of these quantum sensors have achieved state-of-the-art measurements of gravitational acceleration \cite{Menoret2018}, Earth's rotation rate \cite{Gautier2022}, and gravity gradients \cite{Janvier2022}. Current performance levels make these sensors interesting for many different fields. Pushing their precision opens up new horizons, whether in the finer understanding of terrestrial dynamics \cite{Leveque2021} or in exploring the frontier between the quantum and relativistic worlds \cite{Battelier2021}.

Improving the sensitivity of quantum inertial sensors remains a real challenge today. There are two general approaches: (1) reduce the measurement noise (i.e., below the quantum limit), and (2) enlarge the area enclosed by the interferometer, by increasing either the free-fall time $T$ of the atoms or the photon momentum imparted to them \cite{Kovachy2015}. The simplest and perhaps most promising approach is the former, since the sensitivity scales as $T^2$. This is currently the path followed in large-scale infrastructures on ground \cite{Canuel2018, Zhan2019, Asenbaum2020, Badurina2020, Abe2021} and in future Space missions \cite{Elliott2018, Becker2018, Abend2023, Elliott2023}. The Cold Atom Laboratory (CAL) on board the International Space Station and the MAIUS experiment fixed inside a sounding rocket have both demonstrated atom interferometry in Space \cite{Lachmann2021, williams2024interferometry}. On ground, letting atoms fall for longer times necessarily leads to increased apparatus size and complexity. Microgravity facilities, such as the ZARM drop tower \cite{MUNTINGA2013, Raudonis2023} and zero-g aircraft \cite{GEIGER2011, BARRETT2016}, offer very limited up time---making long-term studies challenging. One way of circumventing these limitations is to simulate microgravity conditions using a lab-scale Einstein Elevator (EE). Just as in the \emph{gedanken} experiment envisioned by Einstein in 1907, atoms inside a vacuum chamber are fixed to a moving platform that undergoes near-perfect free fall. Importantly, just as for CAL, the atoms remain accessible to scientists at all times.

In this article, we present the first atom interferometer (AI) measurements on a 3-meter-high EE. This platform provides up to 500 ms of microgravity every 13.5 seconds with a high-degree of reproducibility. It combines the capability of reaching ultra-high sensitivities with performing extended studies suitable for precision measurements. With a repetition rate similar to that of many quantum gas experiments, we perform experiments in microgravity for cumulative times of $> 1$ hour per day---equivalent to more than six 3-hour flights on the Novespace Zero-g Airbus. Operating in microgravity requires the use of a unique light-pulse interrogation scheme [see Fig.~\ref{fig:AtomInterferometryMicrogravity}(b)], which we apply to a sub-Doppler cooled sample of $^{87}$Rb atoms. We perform a Bayesian analysis to extract the fringe contrast and detection noise of our AI and demonstrate a statistical sensitivity of $6 \times 10^{-7}$ m/s$^2$ per shot. We expect this performance to improve with the use of ultra-cold atoms in microgravity \cite{Condon2019}.


\begin{figure}[!ht]
  \centering
  \includegraphics[width=1\textwidth, trim=0mm 0mm 0mm 0mm,, clip]{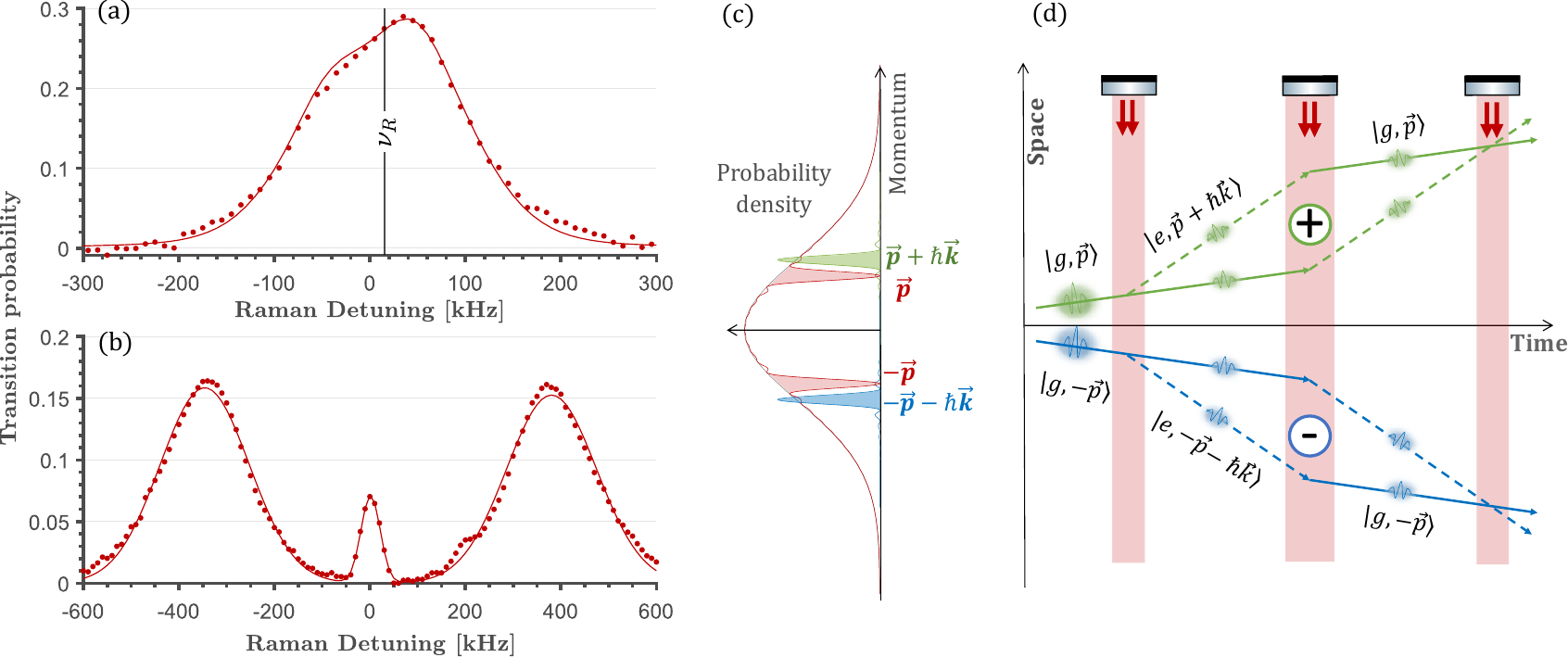}
  \caption{\textbf{Raman spectroscopy in microgravity.} Counter-propagating Raman spectrum of a thermal $^{87}$Rb cloud ($\mathcal{T} = 7.5$ $\mu$K) in microgravity (a) and in standard gravity (b). A $\pi$-pulse pulse is applied after a short time of free fall (7 ms), while scanning the Raman detuning $\delta = \nu_{\rm laser} - \nu_{\rm c}$. The recoil frequency is $\nu_{\rm R} \simeq 15$ kHz. Red points are experimental data while the solid red curve is the simulated spectrum. (c) Velocity distribution of the cold atom cloud and momentum transfer for the two momentum classes $\pm \bm{p}$. (d) Double AI (+) and (-) in the double single diffraction (DSD) regime.}
  \label{fig:Spectro}
\end{figure}

We first investigate the coherent state manipulation of the atoms by performing two-photon Raman spectroscopy in microgravity. Figure \ref{fig:Spectro}(a) depicts the spectrum of a counter-propagating Raman transition for a thermal $^{87}$Rb atom cloud under microgravity conditions. The transition probability from the initial state $\ket{F=1}$ to the final state $\ket{F=2}$ is measured by fluorescence detection while scanning the two-photon Raman detuning $\delta = \nu_1 - \nu_2 - \nu_{\rm c}$, where $\nu_1$ and $\nu_2$ represent the optical Raman frequencies, and $\nu_{\rm c} \simeq 6.834$ GHz is the clock transition frequency of $^{87}$Rb. In standard gravity [Fig.~\ref{fig:Spectro}(b)], the spectrum contains two primary peaks corresponding to two distinct counter-propagating transitions, oppositely frequency shifted due to the Doppler effect, and a resonance at the clock transition indicative of a residual velocity-insensitive co-propagating transition. The width of the counter-propagating peaks provides measurements of the effective two-photon Rabi frequency $\Omega_{\rm eff} = 2\pi\times 15$ kHz, the atom cloud temperature $\mathcal{T} = 7.5$ $\mu$K, and the ratio $\epsilon = \Omega_{\rm co}/\Omega_{\rm eff} = 0.07$, with $\Omega_{\rm co}$ the Rabi frequency of the co-propagating transition. This ratio results from imperfect perpendicular linear polarization between the two Raman beams.

In microgravity [Fig.~\ref{fig:Spectro}(a)], where the Doppler effect vanishes, the spectrum features a single resonance representing the two degenerate counter-propagating Raman transitions. The shape of the measured spectrum is consistent with the results of our simulation, which considers a 10-level system with a residual co-propagating transition due to an imperfectly polarized Raman beam (see Methods). The model incorporates the experimental parameters extracted from the spectrum obtained under standard gravity with no free parameters.

\begin{figure}[!ht]
  \centering
  \includegraphics[width=1\textwidth]{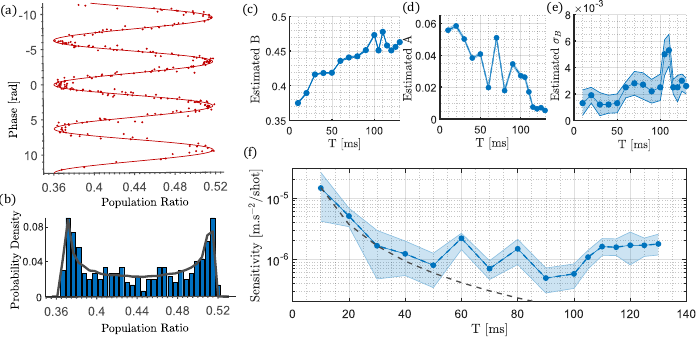}
  \caption{\textbf{Atom interferometry on the Einstein elevator.} (a) Interferometry fringes obtained in the DSD regime ($2T = 20$ ms) and reconstructed via correlation with a classical accelerometer. (b) Histogram (blue) of population ratio measurements from (a) and the probability density function (grey curve) determined by Bayesian analysis. (c), (d) and (e) : Evolution of the parameters with the interrogation time estimated by the Bayesian analysis : offset, contrast and offset noise.  Filled areas show the 95$\%$ confidence interval provided by this analysis. (f) Shot-to-shot sensitivity of the AI as a function of interrogation time $T$. The gray dashed line illustrates the $1/T^2$ scaling of the sensitivity for $T \leq 50$ ms.}
  \label{fig:AcceleroAtomique}
\end{figure}

A Mach-Zehnder AI \cite{KASEVICH1992} is constructed using a $\pi/2-\pi-\pi/2$ sequence of light pulses separated by a free-fall time $T$. In microgravity, our interferometer is in the double single diffraction (DSD) regime \cite{BARRETT2016} where two opposite velocity classes $v_{\rm sel} = \pm 6.8$ mm/s are selected simultaneously---producing two symmetric interferometers [see Fig.~\ref{fig:Spectro}(d)]. Labeling the output of the two respective interferometers $+$ and $-$ for the oppositely directed Raman wavevectors ($\pm \bm{k}_{\rm eff}$), we have:
\begin{equation}
  \frac{N_2(\pm)}{N_{\rm T}(\pm)} = B + A \cos\left(\Phi_{\rm las} \pm \Phi_a \right),
\end{equation}
where $N_2$ is the numbers of atoms detected in $\ket{F=2}$ ground state, $N_{\rm T}$ is the total number of atoms participating in each AI, the fringe amplitudes $A$ and offsets $B$ are assumed to be identical for the two cases, $\Phi_{\rm las}$ is the common laser phase imprinted on the atoms during the AIs, and $\Phi_a = \bm{k}_{\rm eff} \cdot \bm{a} T^2$ is the inertial phase. The fluorescence from both AI outputs are imaged onto a single photodiode and the total population ratio $R$ is obtained using time-resolved fluorescence detection (see Methods). Provided an equal number of atoms participate in both interferometers (i.e., $N_{\rm T}(+) = N_{\rm T}(-)$), the total population ratio can be shown to be:
\begin{equation}
  R = \frac{N_{2}(+) + N_{2}(-)}{N_{\rm T}(+) + N_{\rm T}(-)} = B + A \cos\left(\Phi_{\rm las}\right) \cos\left(\Phi_a\right).
  \label{eq:OutputAIDSD}
\end{equation}
The fringe pattern now contains the product of two sinusoids containing the laser phase and the inertial phase, respectively. Typically, we fix the laser phase at $\Phi_{\rm las} = 0$ to maximize the visibility of the acceleration-sensitive fringe. Figure \ref{fig:AcceleroAtomique}(a) shows interference fringes recorded on the EE using an interrogation time of $2T = 20$ ms. Only one AI measurement is carried out during each parabolic trajectory of the EE, hence 300 such cycles were used to construct the interference pattern. Here, the inertial phase $\Phi_a$ is scanned randomly by the vibrations of the moving platform. Fringes are reconstructed from noise by correlating the output of the AI with a classical accelerometer (see Methods).

When the total interrogation time is greater than $2T = 40$ ms, the self-noise of the classical accelerometer Colybris SF3000L ($\sim 5\times 10^{-6}$ m.s$^{-2}$/Hz$^{1/2}$) dominates the vibration phase estimates and the full interference fringe cannot be reconstructed \cite{BARRETT2015}. Yet amplitude information can still be obtained by building a histogram of measured population ratios $R$. The resulting bimodal probability density [see Fig.~\ref{fig:AcceleroAtomique}(b)] is the sinusoidal response of the AI convolved with the noise distributions of fringe parameters $A$ and $B$. Just as the amplitude and offset uniquely affect the interference fringe, their respective noise distributions produce distinct features in the probability density of $R$. For the purposes of this work, we model the noise spectra for both the amplitude and offset as normal distributions parameterized by mean values $A$ and $B$, and standard deviations $\sigma_A$ and $\sigma_B$, respectively. These parameters are challenging to extract from histograms using heuristic approaches such as non-linear least-squares fits---particularly when the number of measurements is limited. Instead, we apply a novel Bayesian analysis to estimate these noise parameters with a high degree of confidence (see Methods). Assuming the offset noise being the dominant source of noise, allowing us to treat the amplitude $A$ as a constant, we use estimates of $A$ and $\sigma_B$ to determine the shot-to-shot sensitivity $\mathbb{S}$ of the AI:
\begin{equation}
  \mathbb{S} = \frac{\sigma_B}{A \, k_{\rm eff} \, T^2}.
\end{equation}
Figure \ref{fig:AcceleroAtomique}(f) shows the sensitivity of the AI for interrogations times $T$ up to 130 ms. As expected, the sensitivity improves as $T^2$ until $T \simeq 50$ ms. Subsequent increases in $T$ do not produce significant improvements in sensitivity. This behaviour can be explained by a loss of fringe contrast ($2A$) due to cloud expansion in the Raman beam and due to the residual rotation rate of the EE platform. Rotations cause a finite spatial separation of the atomic wavepackets during the last beamsplitter pulse (see Methods)---causing a significant loss of contrast at our relatively high sample temperature (7.5 $\mu$K). The sensitivity reaches $\mathbb{S} = 6 \times 10^{-7}$ m/s$^2$ per shot at $T = 100$ ms---the best AI sensitivity achieved in microgravity so far, corresponding to a 350-fold improvement over previous measurements \cite{GEIGER2011}.


Quantum sensors based on matter-wave interferometry are anticipated to reach their full potential in a microgravity environment. Our results complement other studies in microgravity \cite{GEIGER2011, BARRETT2016, MUNTINGA2013, Raudonis2023, Lachmann2021, williams2024interferometry} that demonstrate the challenges in suppressing vibration and rotation noise, particularly for single-sensor performance. Recent results from the CAL experiment onboard the ISS have demonstrated clear limitations from ambient vibrations \cite{williams2024interferometry}. With differential-sensors utilizing correlated atomic sources \cite{BARRETT2022}, we are able to push the performance further, even in the presence of this noise.

As a case study, we investigate what level of sensitivity our current experiment could achieve with two correlated sources. We consider a test of the Weak Equivalence Principle (WEP) using a simultaneous dual-species interferometers of $^{39}$K and $^{87}$Rb. This cornerstone of General Relativity states that any two objects in the same gravitational field will accelerate at the same rate, regardless of their internal composition. Experiments using chemical species with a large mass difference are expected to be more sensitive to violations of the WEP \cite{SCHLIPPERT2014}. 
These tests rely on precise measurements of the differential acceleration between two test masses. In a dual-species interferometer, this manifests as a differential phase shift $\phi_d$. Extracting the differential phase from the output of two correlated AIs with identical scale factors ($S_j = k_j T^2$ for each species $j$) typically involves ellipse fitting \cite{FOSTER2002} or Bayesian techniques \cite{Stockton2007}. However, if the AI scale factors are different, the AI output forms a Lissajous curve that cannot be fit with the standard approach. In previous work \cite{BARRETT2015}, we developed a generalized Bayesian method to extract $\phi_d$ from such curves and showed that it provides an optimal unbiased estimator of the differential phase. Using simulations, we now investigate the sensitivity of a future WEP test at current noise levels.

\begin{figure}[!ht]
  \centering
  \includegraphics[scale=0.72]{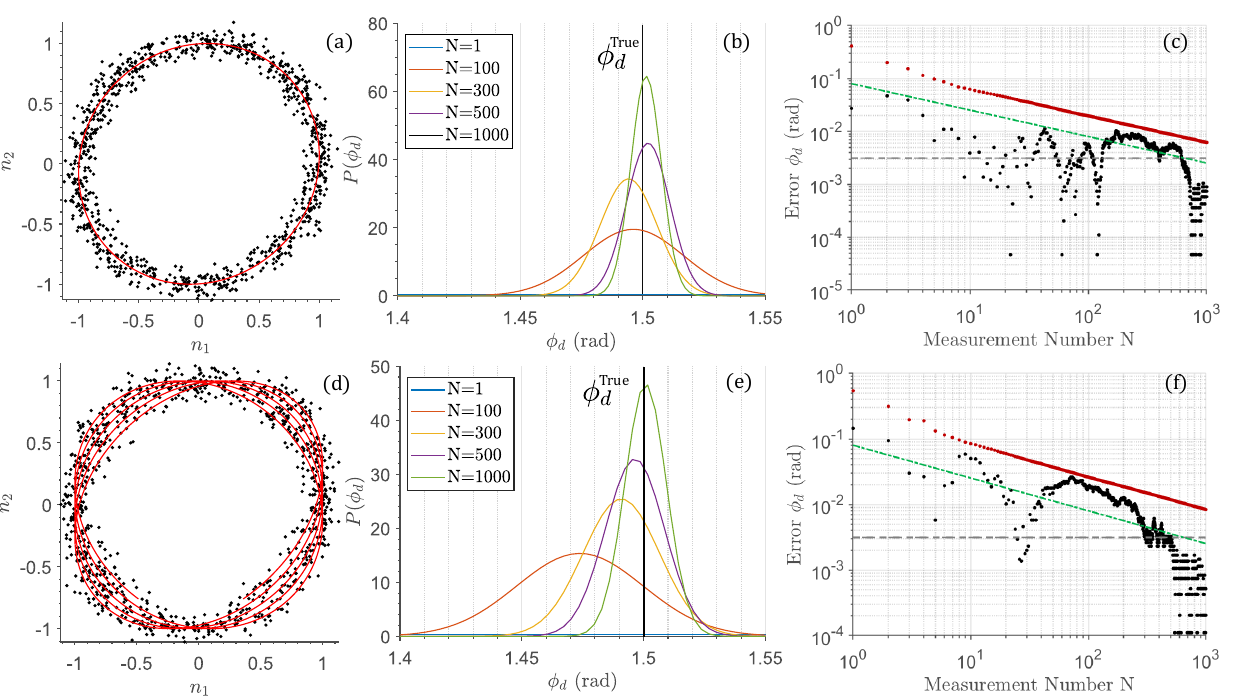}
  \caption{\textbf{Bayesian estimation of differential phase $\phi_d$ in microgravity.} Simulated data for a simultaneous $^{39}$K -- $^{87}$Rb interferometer on the EE platform. Top row: case of Bragg-type interferometers with identical scale factors ($S_{\rm K} = S_{\rm Rb}$). Bottom row: case of Raman-type interferometers with non-identical scale factors ($S_{\rm K} = 1.0177\,S_{\rm Rb}$). (a) and (d): Simulated data (black) and Lissajous curve (red) for total interrogation time $2T = 180$ ms. Here, $N = 1000$ points were generated following the model of Eq.~\eqref{syst3} (in Methods) using $\phi^{\rm true}_d = 1.5$ rad and common phase determined by the level of vibrations on the EE platform. (b) and (e): Conditional probability distributions generated by Bayesian analysis at different $N$. (c) and (f): Statistical uncertainty (red points) and systematic error (black points, discussed in Methods) in the Bayesian estimate of $\phi_d$. To determine these statistics, we perform 10 independent trials for each measurement number $N$. The gray dashed line represents the nominal phase resolution (3 mrad) used in the simulations. The green dashed line indicates the minimum convergence rate in the absence of offset noise using $\sigma_{\phi_d}/\sqrt{N}$.}
  \label{fig:BayesianSimulator}
\end{figure}

Figures \ref{fig:BayesianSimulator}(a) and (d) show simulated data from a $^{39}$K -- $^{87}$Rb interferometer on the EE platform (see Methods for details). In (a), we consider the case of Bragg-type interferometers with identical scale factors \cite{Elliott2023}. Here, a parametric plot of the output from each AI produces an ellipse with an eccentricity determined by $\phi_d$. In (d), we consider Raman-type interferometers with non-identical scale factors and we obtain a Lissajous curve with several loops due to the large phase range spanned by vibrations on the EE platform ($\sim 30$ rad). Figures \ref{fig:BayesianSimulator}(b) and (e) illustrate the progressive improvement in the probability distribution produced by the Bayesian algorithm as more measurements are included. The differential phase estimate is given by the maximum likelihood value of this distribution, and the statistical uncertainty $\Delta \phi_d^{\rm stat}$ is determined by its width. Figures \ref{fig:BayesianSimulator}(c) and (f) show the convergence of the algorithm through a statistical analysis of several independent datasets. Using $N = 1000$ measurements at current noise levels, we reach $\Delta \phi_d^{\rm stat}/\sqrt{N} = 6.5$ mrad for Bragg interferometers (c), and $\Delta \phi_d^{\rm stat}/\sqrt{N} = 8.5$ mrad for Raman interferometers (f). Thus, at current noise levels, these simulations indicate a future WEP test on the EE platform would yield $\Delta \eta^{\rm stat} = \Delta \phi_d^{\rm stat}/\sqrt{N} S_{\rm K} g \simeq 5.1 \times 10^{-9}$ (Bragg) and $6.5 \times 10^{-9}$ (Raman). A survey of 100 datasets would then yield a measurement of $\eta$ at the $10^{-10}$ level.


To conclude, we present the first matter-wave interferometry experiments on an Einstein elevator. We demonstrate an atom accelerometer in microgravity for interrogation times 
up to $2T = 260$ ms, the largest of any experiment to date. Although the coherence of the atomic wavepackets is preserved over the full parabolic trajectory of the elevator (see Methods), the sensitivity to acceleration is currently limited by the contrast loss due to the residual rotation rate of the platform. This can be improved using ultra-cold atom sources, as previously demonstrated on the same apparatus \cite{Condon2019}. We developed a novel Bayesian method to extract interference fringe parameters without the need for a stable interferometer phase. We used this method to demonstrate an acceleration sensitivity of $6 \times 10^{-7}$ m/s$^2$ per shot in microgravity---corresponding to a $\times 350$ improvement over previous cold-atom-based accelerometers in weightlessness \cite{GEIGER2011, BARRETT2016}. Finally, guided by simulations at current noise levels, our platform could yield a test of the Weak Equivalence Principle at the $10^{-10}$ level.

With up to 500 ms of weightlessness every 13.5 seconds, our laboratory-scale Einstein elevator can provide more aggregated time in microgravity ($> 1$ h/day) than any other existing platform on Earth, making it a realistic alternative to atomic fountains \cite{Guena2012}. Its high repetition rate and excellent repeatability are compatible with long-term experimental studies to characterize sensor stability and systematic effects. Our Einstein Elevator has the potential to prepare future Space missions and is an alternative solution to realize experiments in microgravity as currently proposed for the ISS \cite{Aveline2020, Elliott2023}. This paves the way for future metrology experiments such as satellite gravimetry \cite{Leveque2021}, quantum tests of the UFF \cite{Battelier2019}, and gravitational wave detection \cite{DIMOPOULOS2009}.





\newpage{}
\section*{Methods}
\subsection*{Experimental setup and measurement sequence} 


The head consists of a titanium vacuum chamber, ion pump, magnetic gradient and bias compensation coils, a mu-metal magnetic shield, and all optics required for the 3D magneto-optical trap (MOT) and Raman excitation. The vertical Raman beam is retro-reflected by a mirror that acts as an inertial reference frame. A three-axis mechanical accelerometer (Colybris SF3000L) is fixed to the rear of this mirror to monitor its motion. Other critical hardware, including power supplies, RF synthesizers, analog and digital electronics, and fibered laser system, reside in racks near the EE. They are connected to the sensor head through a series of 15-m-long coaxial cables and optical fibers. Additional details can be found in \cite{BARRETT2015, BARRETT2016, Condon2019, BARRETT2022}. 

In this work, the sensor head is fixed to the Einstein Elevator (EE) platform. The EE was designed and built by the French company Symétrie\footnote{Based in Nîmes, France, the company develops hexapod systems and high-precision positioning.}. Its operational principle relies on mimicking the free-fall trajectory of an object in Earth's gravity. The sensor head, fixed to its moving platform, experiences up to half a second of weightlessness by launching it upward and following a pre-programmed parabolic trajectory using a precisely-controlled motorized position servo. Measurements of the moving platform's position are provided by two incremental grating rulers. Vertical accelerations are achieved thanks to two linear motors located on both sides of the platform. The cycling rate of the EE is limited by the 12 s cool-down period required by the linear motors.

\begin{figure}[!ht]
  \centering
  \includegraphics[width=0.9\textwidth]{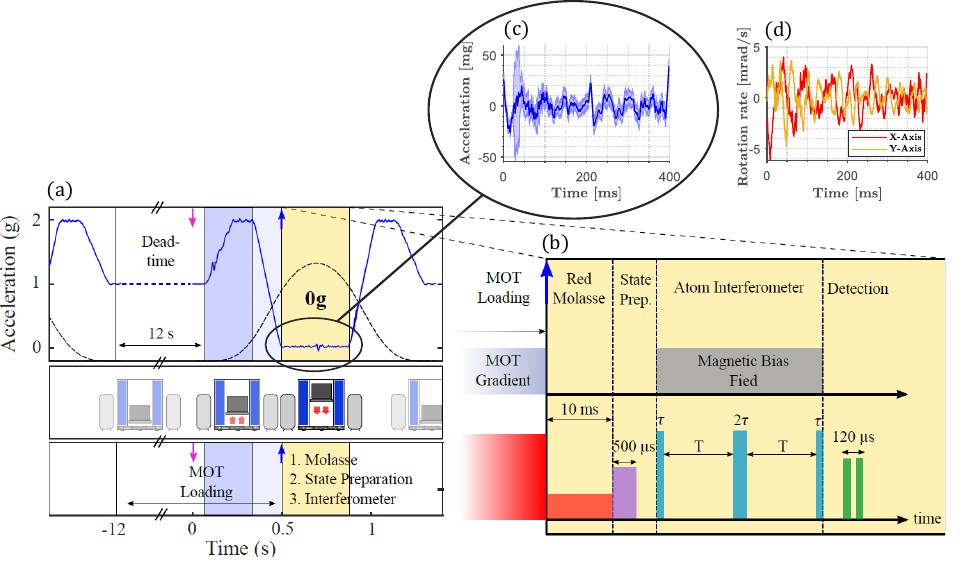}
  \caption{\textbf{Sequence of Atom interferometry on the Einstein Elevator:} (a) Acceleration (solid blue curve) and position (dashed black curve) profiles of an platform trajectory with $t = 400$ ms in 0g. (b) Cold atom interferometer sequence. The MOT loading starts a few seconds before the start of the motion. The molasses cooling, state preparation, Raman interferometer, and detection are performed during the 0g phase, after receiving a trigger from the EE controller. (c) Mean residual acceleration along the vertical axis during the 0g phase from 300 parabolic trajectories. The shaded regions indicate the standard deviation of those data. (d) Typical residual rotation rates about the horizontal X- and Y-axes ($\Omega_{\rm X}$, $\Omega_{\rm Y}$) during the 0g phase.}
  \label{fig:sequenceS0g}
\end{figure}

We evaluated the quality of the EE trajectory by recording the acceleration of the reference mirror and the rotation rate of the platform during its motion. Figure \ref{fig:sequenceS0g}(a) shows the typical acceleration profiles obtained throughout one full cycle. The vertical acceleration profile is comprised of standard gravity (1g), hypergravity (2g), and microgravity (0g) phases. Figure \ref{fig:sequenceS0g}(c) shows residual vibrations measured by the mechanical accelerometer during the 0g phase. The peak-to-peak residual acceleration is below 50 mg on the vertical Z-axis and below 100 mg on the horizontal X- and Y-axes (not shown). Rotation rates about the horizontal axes ($\Omega_{\rm X}$ and $\Omega_{\rm Y}$) are monitored using two fiber optic gyroscopes (KVH DSP-1750) fixed directly to the elevator platform. Figure \ref{fig:sequenceS0g}(d) shows a typical rotation rate profile during the 0g phase, which features peak-to-peak values less than 10 mrad/s.

The experimental sequence is illustrated in Fig.~\ref{fig:sequenceS0g}(b). A six-beam MOT of $^{87}$Rb is loaded directly from background vapor for typically 4 seconds with an optical power of about 100 mW ($20 \, I_{\rm sat}$). The cooling beams are red-detuned by $-18$ MHz ($-3\,\Gamma$) from the $^{87}$Rb cycling transition and the ratio of cooling to repump light intensity is $I_{\rm rep}/I_{\rm cool} = 0.1$. About $5 \times 10^7$ atoms are loaded in this configuration. Atoms are fully loaded into the MOT during the dead time between parabolas, and they are maintained during the pull-up and injection phases. When the EE platform reaches the 0g-phase, it triggers our control system to initiate the sub-Doppler cooling stage. This stage is performed during the first 10 ms of the 0g phase, where the MOT gradient coils are turned off, the detuning of the cooling beams is linearly ramped to $-24\,\Gamma$, and the overall cooling power is reduced to $\sim 1\,I_{\rm sat}$. At this point the cloud temperature reaches $\sim 7$ $\mu$K with all ground state magnetic sub-levels populated. The atoms are then prepared in the $\ket{F=1, m_F=0}$ state via a purification sequence. First, a ``repump'' pulse resonance with the $\ket{F = 1} \to \ket{F'=2}$ transition pumps atoms to the $\ket{F = 2}$ state. Then, a magnetic field bias of 130 mG is applied along the vertical direction to split the magnetic sub-levels. This is followed by a microwave $\pi$-pulse, resonant with the clock transition at 6.834 GHz, 500 $\mu$s in duration that transfers population from $\ket{F = 2, m_F = 0}$ to $\ket{F = 1, m_F = 0}$. Finally, we remove atoms remaining in the $\ket{F=2}$ manifold with a 600 $\mu$s push pulse resonant with the $\ket{F = 2} \to \ket{F' = 3}$ cycling transition. Only atoms in the $\ket{F=1, m_F=0}$ remain. This state preparation sequence is followed by a sequence of Raman pulses that form the atom interferometer. A $\pi/2$ -- $\pi/2$ sequence is used to perform optical Ramsey spectroscopy to determine the coherence of the sample at large free-fall times (see below). We use a $\pi/2$ -- $\pi$ -- $\pi/2$ pulse sequence to construct the quantum accelerometer. The detuning of the Raman beam from the $\ket{F' = 2}$ excited state is $\Delta \simeq -830$ MHz and the duration of the Raman $\pi$-pulse is $2\tau \simeq 20$ $\mu$s. State detection is performed by collecting atomic fluorescence on a photodiode. We apply two light pulses to the sample after the interferometer along the vertical direction---one resonant with the $\ket{F=2} \rightarrow \ket{F'=3}$ transition to measure the number of atoms in $\ket{F = 2}$ ($N_2$), and one that includes an additional repump sideband, giving the total number of atoms in both ground states ($N_{\rm T}$). The peak intensity of these pulses is $\sim 37 \, I_{\rm sat}$.

To reconstruct the blurred atomic fringes due to the residual vibrations of the EE platform, we make use of the mechanical accelerometer fixed to the rear of reference mirror, measuring the acceleration perpendicular to the mirror's surface. During the interferometer, this time-varying acceleration is weighted by the response function of the AI \cite{Cheinet2008}, and integrated over the total interrogation time---providing an estimate of the vibration-induced phase shift. AI fringes are reconstructed from noise by plotting the measured population ratio as a function of this estimated phase shift \cite{BARRETT2015}.

\subsection*{Ramsey interferometry} 

To assess our ability to manipulate atoms throughout the entirety of the zero gravity phase provided by our simulator, we conduct Ramsey fringes. Ramsey interferometry is widely used for high-resolution spectroscopy, primary frequency standards in cesium fountain clocks \cite{Gibble2016}, and tests of fundamental physics with optical atomic clocks \cite{Ludlow2015}. The frequency sensitivity of the Ramsey technique scales inversely with the interrogation time $T_{\rm R}$ in a $\pi/2-\pi/2$ sequence of pulses. Recent experiments have pushed the limits of Ramsey interrogation in reduced gravity environments \cite{Liu2018}, achieving up to $T_{\rm R} = 400$ ms on a 0g aircraft \cite{Langlois2018}. Here, we probe the $^{87}$Rb clock transition on the EE using co-propagating optical Raman beams with interrogation times up to 380 ms.

\begin{figure}[!th]
  \centering
  \includegraphics[width=0.9\textwidth]{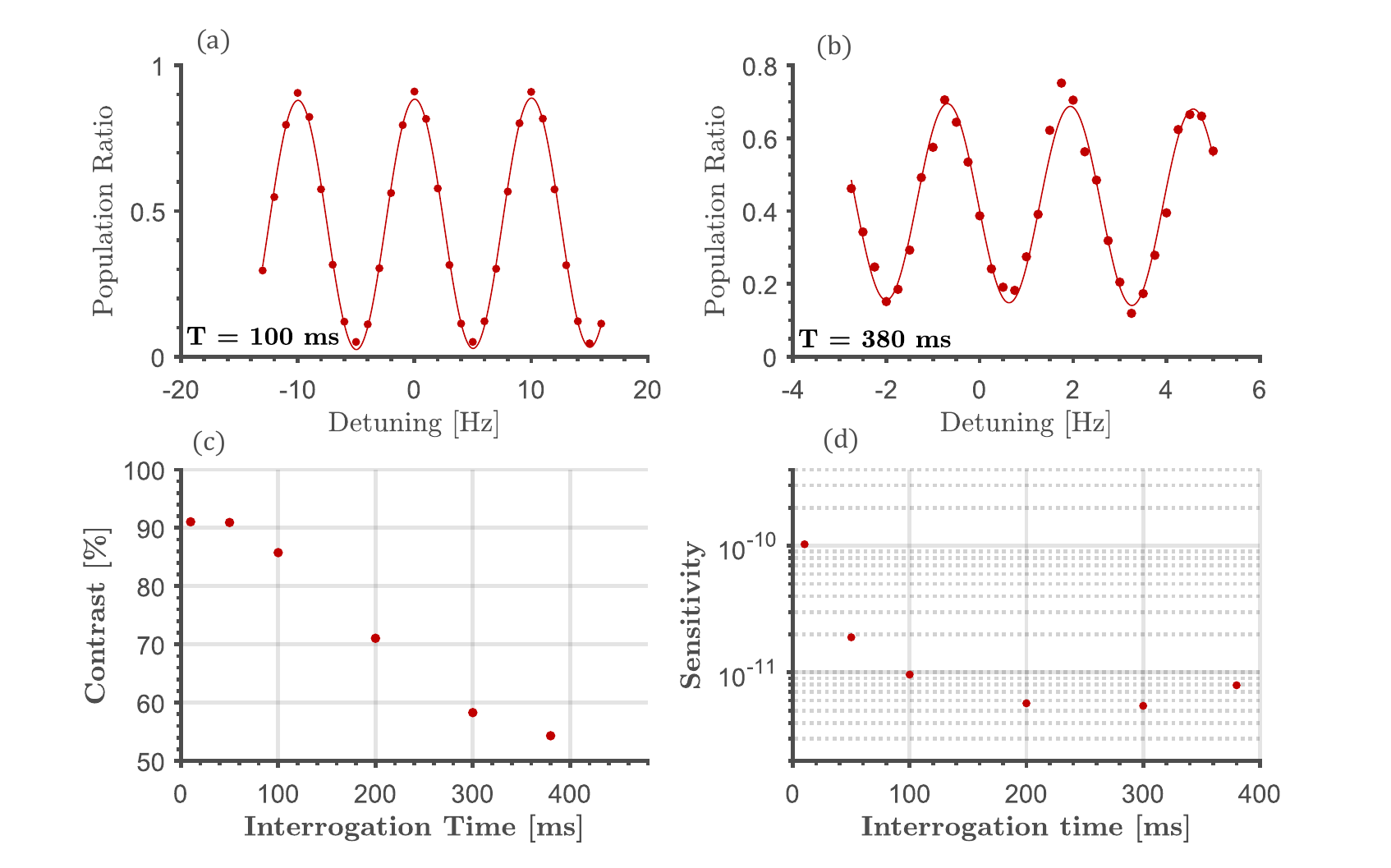}
  \caption{\textbf{Ramsey interferometry.} (a) and (b) Ramsey fringes in microgravity for an interrogation time of respectively $T_{\rm R}=100$ ms and $T_{\rm R}=380$ ms. (c) Contrast of the Ramsey fringes and (d) short term sensitivity on the measured frequency as a function of the interrogation time $T_{\rm R}$.}
  \label{fig:Ramsey}
\end{figure}

Figure \ref{fig:Ramsey}(a) and (b) shows Ramsey fringes obtained for $T_{\rm R} = 100$ ms and $380$ ms. Fits to these data yield a fringe half-width at half-maximum (HWHM) of $\Delta \nu = 1/2T_{\rm R}$ as low as 1 Hz at $T_{\rm R} = 380$ ms. The uncertainty on the fit of the fringes is less than 1 Hz, and the lightshift is not corrected in these datasets. This measurement is not conducted for metrological purposes, but it illustrates the potential for exploring long-time interference effects in our microgravity simulator. Figure \ref{fig:Ramsey}(d) shows measurements of the short-term frequency sensitivity $\sigma_{\nu}$ as a function of the interrogation time. Analytically, $\sigma_{\nu}$ is given by the following expression:
\begin{equation}
  \sigma_{\nu} = \frac{1}{\pi}\frac{\Delta \nu}{\nu_c}\frac{1}{\text{SNR}}, 
\end{equation}
where ${\rm SNR} = C/2\sigma_{\rm res}$ is the signal-to-noise ratio of the Ramsey fringes determined by the fringe contrast $C$ and standard deviation of fit residuals $\sigma_{\rm res}$. The measured fringe contrast decreases with the interrogation time, as shown in Figure ~\ref{fig:Ramsey}(c). This is due to the finite temperature of the atoms: as the cloud expands, it samples the lower-intensity regions of the Gaussian Raman beam, which reduces the effective Rabi frequency $\Omega_{\rm co}$ and therefore the efficiency of the final beamsplitter pulse. As a consequence, the short-term sensitivity of the measurement is limited to $\sigma_{\nu} = 8 \times 10^{-12}$ for $T_{\rm R} = 380$ ms. Our best sensitivity $\sigma_{\nu} = 5.4 \times 10^{-12}$  is reached for $T_{\rm R} = 300$ ms. This corresponds to a trade-off between the gain with interrogation time and the reduction of fringe contrast.

\subsection*{Double Diffraction and Double Single Diffraction} 

The two optical frequencies required to drive a two-photon Raman transitions are derived from an optical phase modulator. They are therefore phase-locked, spatially overlapped, and feature the same optical polarization. We inject this light into a single-mode polarization-maintaining fiber connected to a commercial beam collimator that expands the Raman beam diameter to $\sim 20$ mm. This beam is aligned vertically through the atoms, with one $\lambda/4$-plate on either side of the vacuum system, before being retro-reflected by the reference mirror. With a suitable choice of the first $\lambda/4$-plate angle, the polarization of the two Raman beams will be lin$\perp$lin, where they drive velocity-sensitive transitions (counter-propagating configuration), or $\sigma^+/\sigma^-$, where they drive velocity-insensitive transitions (co-propagating configuration). In the latter case, the two Raman transitions are degenerate due to the absence of a Doppler shift. This transition is visible near zero detuning in Fig.~\ref{fig:AtomInterferometryMicrogravity}(b) due to imperfect lin$\perp$lin polarization. Under gravity, accelerated atoms experience a Doppler shift that lifts the degeneracy of the two counter-propagating transitions. In weightlessness, all of these Raman transitions are degenerate and occur simultaneously.

Because our thermal cloud has a relatively large range of velocities ($\gg \hbar k_{\rm eff}/M$ where $M$ is the mass of one atom of Rubidium), two regimes of atomic diffraction are possible. For the zero-momentum class, $\ket{F=1,p=0}$ is simultaneously coupled with both states $\ket{F=2, p=\pm \hbar k_{\rm eff}}$ leading to double diffraction \cite{LEVEQUE2009}. For a momentum class $p \neq 0$, the laser pulse couples simultaneously $\ket{F=1, p}$ with $\ket{F=2, p+\hbar k_{\rm eff}}$ on one hand, and $\ket{F=1, -p}$ with $\ket{F=2, -p-\hbar k_{\rm eff}}$ on the other hand---leading to double single diffraction \cite{BARRETT2016}. The double diffraction regime is avoided by choosing Rabi frequency $\Omega_{\rm eff}/2\pi \simeq \nu_{\rm R}$, which determines the pulse bandwidth, and a Raman detuning $|\delta| > \nu_{\rm R}$, where $\nu_{\rm R} \simeq 15$ kHz is the recoil frequency [see Fig.~\ref{fig:AcceleroAtomique}(a)]. Typically, we use $\delta = 50$ kHz. Under these conditions, the first Raman pulse selects a narrow range of atomic velocities symmetrically about $p = 0$. The double single diffraction process leads to two symmetric AIs as discussed above [see Fig.~\ref{fig:AcceleroAtomique}(b)]. Only a fraction of the total number of atoms participate in the upper ($+$) or the lower ($-$) interferometers. 

After the interferometer sequence, position-velocity correlation in the two diffracted wavepackets leads to two spatially-separated clouds. For $2T = 200$ ms, the two interferometer ports are separated by 8 mm, which is small compared to the width of each cloud ($\sim 20$ mm) due to finite temperature of the sample. We detect the fluorescence of both clouds simultaneously by imaging the light on a single photodiode. This provides a measurement of the population ratio $R$, and hence the acceleration $\Phi_a$ through Eq.~\eqref{eq:OutputAIDSD}.

\subsection*{Model of Raman Spectroscopy in microgravity} 

\begin{figure}[!t]
  \centering
  \includegraphics[width=1.1\textwidth]{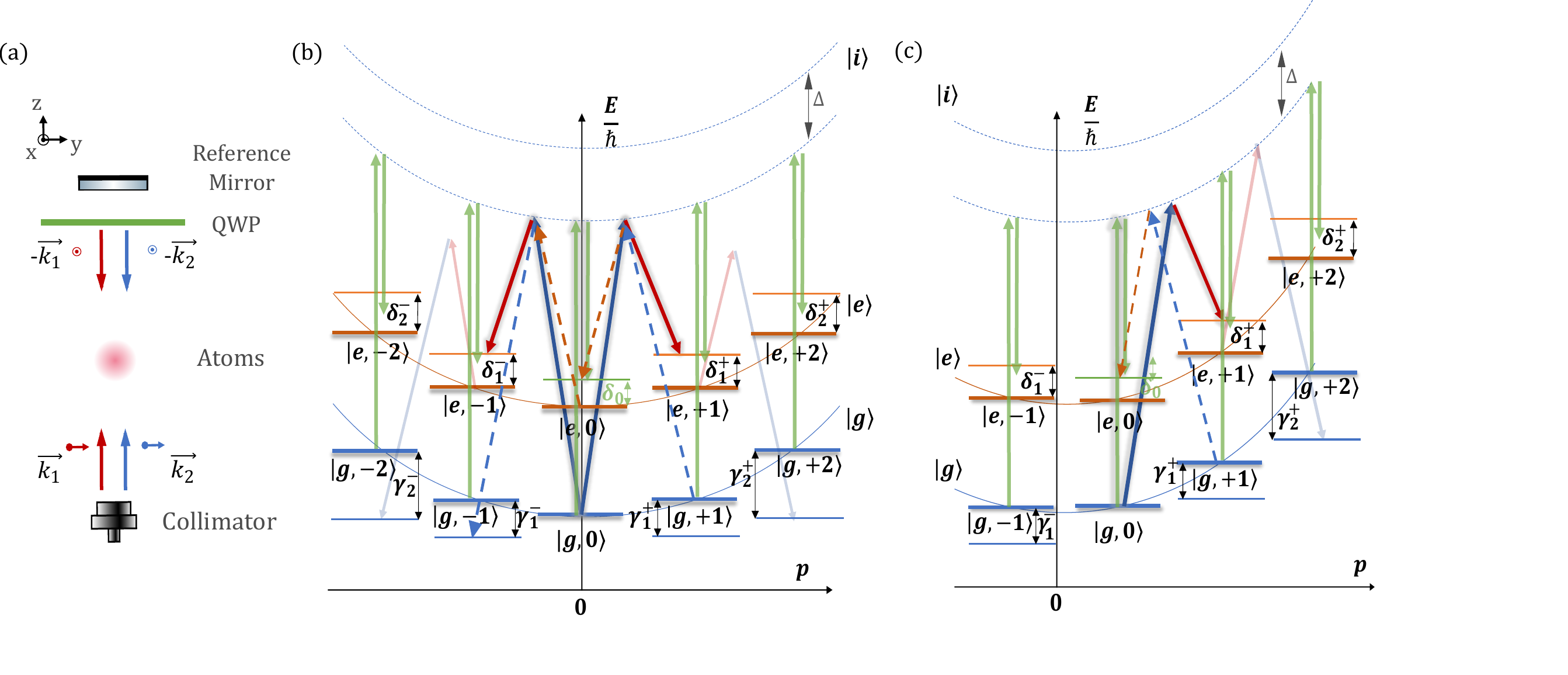}
  \caption{\textbf{Atom-light system considered in the model.} (a) Retro-reflected configuration of the Raman beam. (b) Energy-momentum diagram for momenta in the cloud centered at \( p = 0 \). We include 10 bare states that are coupled to each other by several two-photon processes through an intermediate state \(\ket{i}\). Counter-propagating Raman transitions are represented by pairs of blue (\(\pm \bm{k}_2\)) and red (\(\pm \bm{k}_1\)) arrows. Co-propagating transitions are shown as green arrows. By selecting the zero-momentum class, \(\ket{F=1,p=0}\) ($\ket{g,0}$ in the diagram) is simultaneously coupled with both states \(\ket{F=2, p=\pm \hbar k_{\rm eff}}\) ($\ket{g,\pm 1}$), leading to double diffraction. (c) Energy-momentum diagram for a selected momentum class \(p \neq 0\). The diagram shows the laser pulse coupling \(\ket{F=1, p}\) ($\ket{g,0}$) with \(\ket{F=2, p+\hbar k_{\rm eff}}\) ($\ket{g,+1}$). Simultaneously, but not shown in the diagram for readability, it couples \(\ket{F=1, -p}\) with \(\ket{F=2, -p-\hbar k_{\rm eff}}\), leading to double single diffraction.}
  \label{fig:10level}
\end{figure}

We perform a numerical simulation to model the Raman spectrum in microgravity using measured experimental parameters.
The atom-light interaction leads to a four-photon process that drives Raman transitions with effective wave vectors $\pm \bm{k}_{\rm eff} = \pm(\bm{k}_1 - \bm{k}_2) \simeq \pm 2\hbar \bm{k}_1$, since $\bm{k}_2 \simeq -\bm{k}_1$. The two optical frequency components of the incident beam ($\omega_1$, $\bm{k}_1$ and $\omega_2$, $\bm{k}_2$) have the same linear polarization. Generally, the polarization also contains a small residual circular component that allows co-propagating Raman transitions to contribute to the Raman spectrum. We consider a one-dimensional system of atoms with two long-lived electronic ground states $\ket{e}$ and $\ket{g}$ separated in frequency by $\omega_{\ket{e}} - \omega_{\ket{g}}$. This splitting is much larger than the Doppler width of the sample. Each atom has an arbitrary momentum class $p$ and can be excited by integer multiples of the photon momentum $\hbar k_{\rm eff}$---creating a ladder of coupled momentum states. We truncate this state space to $\pm 2 \hbar k_{\rm eff}$ in both ground states, resulting in a 10-level model of the atom. We label these bare states by $\ket{g,j}$ and $\ket{e,j}$, where $j = 0, \pm 1, \pm 2$ is an integer representing momentum state $p + j \hbar k_{\rm eff}$. The full atom-light system is represented by a energy-momentum diagram shown in Fig.~\ref{fig:10level}.

We express the time-dependent atomic wavefunction $\ket{\psi(t)}$ as the following superposition of bare states $\ket{n}$ with corresponding amplitudes $c_{\ket{n}}(t)$:
\begin{equation}
  \ket{\psi(t)} = \sum_n c_{\ket{n}}(t) e^{-i \omega_{\ket{n}} t} \ket{n}.
\end{equation}
Here, the wavefunction is written in the interaction representation, where the basis states rotate at a frequency corresponding to their internal energies $\hbar \omega_{\ket{n}}$. Following the calculation in Ref.~\cite{leveque2010}, we modify the basis to the field-interaction representation. Here, the states rotate at an additional frequency corresponding to the detuning relative the two-photon resonance between the central states $\ket{g,0}$ and $\ket{e,0}$:
\begin{subequations}
\label{chgtvar}
\begin{align}
  a_{\ket{e,-2}}(t) & = c_{\ket{e,-2}}(t) e^{i(\delta_2^{-}+\gamma^{-}_1+\delta_0) t} \\
  a_{\ket{g,-2}}(t) & = c_{\ket{g,-2}}(t) e^{i(\gamma^{-}_2+\delta^{-}_1) t} \\
  a_{\ket{e,-1}}(t) & = c_{\ket{e,-1}}(t) e^{i\delta^{-}_1 t} \\
  a_{\ket{g,-1}}(t) & = c_{\ket{g,-1}}(t) e^{i(\gamma^{-}_1+\delta_0) t} \\
  a_{\ket{g,0}}(t) & = c_{\ket{g,0}}(t) \\
  a_{\ket{e,0}}(t) & = c_{\ket{e,0}}(t) e^{i\delta_0 t} \\
  a_{\ket{g,+1}}(t) & = c_{\ket{g,+1}}(t) e^{i(\gamma^{+}_1+\delta_0) t} \\
  a_{\ket{e,+1}}(t) & = c_{\ket{e,+1}}(t) e^{i\delta^{+}_1 t} \\
  a_{\ket{g,+2}}(t) & = c_{\ket{g,+2}}(t) e^{i(\gamma^{+}_2+\delta^{+}_1) t} \\
  a_{\ket{e,+2}}(t) & = c_{\ket{e,+2}}(t) e^{i(\delta_2^{+}+\gamma^{+}_1+\delta_0) t}
\end{align}
\end{subequations}
where the detunings are as follows:
%
%
\begin{subequations}
\label{eq:Detuning}
\begin{align}
  \delta_0 & = (\omega_1-\omega_2) - (\omega_{e}-\omega_{g}) \\
  \delta_1^+ & = \delta_0 - (\omega_{\rm R} + \omega_{\rm D}) \\
  \delta_1^- & = \delta_0 - (\omega_{\rm R} - \omega_{\rm D}) \\
  \delta_2^+ & = \delta_0 - (3\omega_{\rm R} + \omega_{\rm D}) \\
  \delta_2^- & = \delta_0 - (3\omega_{\rm R} - \omega_{\rm D}) \\
  \gamma_1^+ & = -\delta_2^+ - (4\omega_{\rm R} + 2\omega_{\rm D}) \\
  \gamma_1^- & = -\delta_2^- - (4\omega_{\rm R} - 2\omega_{\rm D}) \\
  \gamma_2^+ & = -\delta_1^+ - (4\omega_{\rm R} + 2\omega_{\rm D}) \\
  \gamma_2^- & = -\delta_1^- - (4\omega_{\rm R} - 2\omega_{\rm D})
\end{align}
\end{subequations}
Here, $\omega_{\rm D} = k_{\rm eff} p/M$ is the Doppler frequency, and $\omega_{\rm R} = \hbar k_{\rm eff}^2/2M$ the two-photon recoil frequency.

The wavefunction in the field-interaction representation is $\ket{\tilde{\psi}(t)} = \sum_n a_{\ket{n}}(t) \ket{n}$. The benefit of this representation is that the effective Hamiltonian becomes time-independent and the dynamics are less computationally costly to determine. The time evolution of the system is computed by numerically solving the Schr\"{o}dinger equation \cite{MOLER1992,LEVEQUE2009}:
\begin{equation}
  i\hbar\frac{d}{dt}\ket{\tilde{\psi}(t)} = \mathbb{H}_{\rm eff} \ket{\tilde{\psi}(t)},
  \label{schro}
\end{equation}
by computing the matrix exponential of the effective Hamiltonian $\mathbb{H}_{\rm eff}$:
\begin{equation}
  \label{eq:TDSE_Soln}
  \ket{\tilde{\psi}(t)} = \exp\left(-\frac{i}{\hbar} \, \mathbb{H}_{\rm eff} \, t \right) \ket{\tilde{\psi(0)}}.
\end{equation}
The effective Hamiltonian can be written as the sum of two matrices:
\begin{equation}
  \mathbb{H}_{\rm eff} = -\hbar (\mathbb{D} + \mathbb{A}),
  \label{Eq:H}
\end{equation}
where $\mathbb{D}$ is a diagonal matrix containing the state-dependent detunings:
\begin{equation}
  \mathbb{D} = \mbox{diag}
  \begin{bmatrix}
    -(\delta_2^- + \gamma_1^- + \delta_0) & = & 4\omega_{\rm R} - 2\omega_{\rm D} - \delta_0 \\
    -(\gamma_2^- + \delta_1^-) & = & 4\omega_{\rm R} - 2\omega_{\rm D} \\
    -\delta_1^- & = & \omega_{\rm R} - \omega_{\rm D} - \delta_0 \\
    -(\gamma_1^- + \delta_0) & = & \omega_{\rm R} - \omega_{\rm D} \\
    & 0 & \\
    & -\delta_0 & \\
    -(\gamma_1^+ + \delta_0) & = & \omega_{\rm R} + \omega_{\rm D} \\
    -\delta_1^+ & = & \omega_{\rm R} + \omega_{\rm D} - \delta_0 \\
    -(\gamma_2^+ + \delta_1^+) & = & 4\omega_{\rm R} + 2\omega_{\rm D} \\
    -(\delta_2^+ + \gamma_1^+ + \delta_0) & = & 4\omega_{\rm R} + 2\omega_{\rm D} - \delta_0 \\
  \end{bmatrix},
\end{equation}
and $\mathbb{A}$ is a matrix with only off-diagonal elements that describe the coupling between states:
{\small
$$
\mathbb{A} = 
\begin{array}{c|cccccccccc} 
           & \ket{e,-2} & \ket{g,-2} & \ket{e,-1} & \ket{g,-1} & \ket{g,0} & \ket{e,0} & \ket{g,+1} & \ket{e,+1} & \ket{g,+2} & \ket{e,+2} \\
\hline \hline 
\bra{e,-2} & 0 & \chi_{\rm co}^* & 0 & \chi_{\rm eff}^* & 0 & 0 & 0 & 0 & 0 & 0 \\
\hline 
\bra{g,-2} & \chi_{\rm co} & 0 & \chi_{\rm eff}^* & 0 & 0 & 0 & 0 & 0 & 0 & 0 \\
\hline
\bra{e,-1} & 0 & \chi_{\rm eff} & 0 & \chi_{\rm co}^* & \chi_{\rm eff}^* & 0 & 0 & 0 & 0 & 0 \\
\hline
\bra{g,-1} & \chi_{\rm eff} & 0 & \chi_{\rm co} & 0 & 0 & \chi_{\rm eff}^* & 0 & 0 & 0 & 0 \\
\hline
\bra{g,0} & 0 & 0 & \chi_{\rm eff} & 0 & 0 & \chi_{\rm co}^* & 0 & \chi_{\rm eff}^* & 0 & 0 \\
\hline 
\bra{e,0} & 0 & 0 & 0 & \chi_{\rm eff} & \chi_{\rm co} & 0 & \chi_{\rm eff}^* & 0 & 0 & 0 \\
\hline 
\bra{g,+1} & 0 & 0 & 0 & 0 & 0 & \chi_{\rm eff} & 0 & \chi_{\rm co}^* & 0 & \chi_{\rm eff}^* \\
\hline 
\bra{e,+1} & 0 & 0 & 0 & 0 & \chi_{\rm eff} & 0 & \chi_{\rm co} & 0 & \chi_{\rm eff}^* & 0 \\
\hline
\bra{g,+2} & 0 & 0 & 0 & 0 & 0 & 0 & 0 & \chi_{\rm eff} & 0 & \chi_{\rm co}^* \\
\hline 
\bra{e,+2} & 0 & 0 & 0 & 0 & 0 & 0 & \chi_{\rm eff} & 0 & \chi_{\rm co} & 0 \\
\end{array},
$$
}
where $\chi_{\rm eff} \equiv \Omega_{\rm eff}/2$ and $\chi_{\rm co} \equiv \Omega_{\rm co}/2$ are half-Rabi frequencies. The effective two-photon Rabi frequency is given by $\Omega_{\rm eff} = \frac{\Omega^*_{gi}\Omega_{ei}}{2\Delta}$, where $\Omega_{gi}$ and $\Omega_{ei}$ are Rabi frequencies associated with the one-photon transitions between $\ket{g}$ and $\ket{i}$, and between $\ket{e}$ and $\ket{i}$, respectively. We account for residual co-propagating transitions with the Rabi frequency $\Omega_{\rm co} = \epsilon \Omega_{\rm eff}$, where $\epsilon \ll 1$ is a factor that describes the defect in linear polarization. These transitions are insensitive to the Doppler effect and transfer negligible recoil to the atoms since $\bm{k}_1 + \bm{k}_2 \simeq 0$. Consequently, co-propagating transitions occur at a fixed detuning $\delta_0$ for all momentum states under consideration.

The simulation takes into account the velocity dispersion of the atomic cloud through the Doppler shift $\omega_{\rm D}(v) = k_{\rm eff} v$ of velocity-dependent transitions. We assume a 1D Maxwell-Boltzmann velocity distribution at temperature $\mathcal{T}$ with probability density:
\begin{equation}
  g(v) = \frac{1}{\sqrt{\pi \sigma_v^2}} \exp\left( -\frac{v^2}{\sigma_v^2} \right),
\end{equation}
where $\sigma_v = \sqrt{2 k_{\rm B} \mathcal{T}/M}$. To compute the population $P_{\ket{n}}(\delta,t)$ of state $\ket{n}$ as a function of laser detuning $\delta$, we integrate our numerical solution \eqref{eq:TDSE_Soln} over this velocity distribution:
\begin{equation}
  P_{\ket{n}}(\delta, t)
  = \int g(v) \left| \braket{n|\tilde{\psi}(v,\delta,t)} \right|^2 dv
  = \int g(v) \left| a_{\ket{n}}(v,\delta,t) \right|^2 dv.
\end{equation}
The total population in the upper ground state $\ket{e}$ is obtained by summing the populations of the five associated momentum states: $\ket{e,-2}, \ket{e,-1}, \ket{e,0}, \ket{e,+1}, \ket{e,+2}$. This model of Raman spectra is shown in Fig.~\ref{fig:Spectro}.

\subsection*{Contrast loss due to rotation noise} 

The sensitivity of the atom interferometer at long interrogation times ($T > 50$ ms) is primarily limited by two effects: (1) reduction of the Rabi frequency due to cloud expansion and spatial averaging of the Raman beam, and (2) the rotation rate of the EE platform and its associated noise. The relatively high temperature of our thermal samples ($\mathcal{T} \approx 7$ $\mu$K) exacerbates both of these effects. The former effect is well understood and has been studied elsewhere \cite{Templier2022}. 
In this section, we describe the latter effect since it has unique features on the EE platform. Rotations of the Raman wavevector cause a separation of atomic trajectories at the final $\pi/2$-pulse. This ``open interferometer'' effect can be quantified by an overlap integral between the wavepackets at either output port of the interferometer. Following previous work \cite{ROURA2014, BARRETT2016}, we introduce the phase space displacement vector:
\begin{equation}
  \delta\bm{\chi}(t) = \delta\bm{P}(t) - \frac{M}{\Delta t} \delta\bm{R}(t).
  \label{eq:SplittingPhaseSpace}
\end{equation}
Here, $\delta\bm{R}(t)$ and $\delta\bm{P}(t)$ are the position and momentum displacement vectors between two atomic wavepackets at time $t$, and $\Delta t$ is the expansion time of the wave function. We consider the two wavepacket trajectories associated with a vertically-oriented, single-diffraction Mach-Zehnder interferometer ($\bm{k}_{\rm eff} = k_z \hat{\bm{z}}$) in a reference frame undergoing a constant rotation with rotation vector $\bm{\Omega} = (\Omega_x, \Omega_y, \Omega_z)$. The phase space displacement at time $t = \Delta t = 2T$ can be shown to be (up to order $T^4$ and $\Omega^2$):
\begin{equation}
  \delta\bm{\chi}(2T) = \frac{\hbar k_z}{2}
  \begin{pmatrix}
    2 \Omega_y T - \Omega_x \Omega_z T^2 \\
    -2 \Omega_x T - \Omega_y \Omega_z T^2 \\
    (\Omega_y^2 + \Omega_z^2) T^2 \\
  \end{pmatrix}.
  \label{eq:DeltaChiConstantRotation}
\end{equation}
The contrast loss is then computed from the following overlap integral
\begin{equation}
  C(t) = \left| \int \exp \left( \frac{i \delta \bm{\chi}(t) \cdot \bm{r}}{\hbar} \right) \left| \psi(\bm{r},t) \right|^2 \rm d^3 \bm{r} \right|
  \label{eq:ContrastLossCalcul}
\end{equation}
where $|\psi(\bm{r},t)|^2$ is the spatial probability density at time $t$. Assuming a thermal velocity distribution, this probability density can be written:
\begin{equation}
  |\psi(\bm{r}, t)|^2 = \frac{1}{\left(\pi \sigma_r^2(t) \right)^{3/2}} \exp\left(- \frac{r^2}{\sigma_r^2(t)} \right)
\end{equation}
where $\sigma_r^2(t) = \sigma_0^2 + \sigma_v^2 t^2$ is the $e^{-1}$ spatial width of the distribution at time $t$, $\sigma_0$ is the initial wavepacket spread, and $\sigma_v = \sqrt{2 k_{\rm B} \mathcal{T}/M}$ is the velocity dispersion. Evaluating the contrast loss up to order $T^4$ and $\Omega^2$, we find:
\begin{equation}
  C(2T)
  \approx \exp\left[-\left(\frac{k_z \sigma_r(2T)}{2} \right)^2 \left( \Omega_x^2 + \Omega_y^2 \right) T^2 \right]
  \approx \exp\left[-k_z^2 \sigma_v^2 \left( \Omega_x^2 + \Omega_y^2 \right) T^4 \right],
  \label{eq:ContrastLoss}
\end{equation}
where we made the approximation $\sigma_r(2T) \approx 2 \sigma_v T \gg \sigma_0$.

For the more general case of a time-varying rotation rate $\bm{\Omega}(t)$, we numerically compute the orientation of the Raman beam using a two-axis gyroscope mounted directly on the EE platform. We apply a 3D rotation matrix that orients the effective wavevector $\bm{k}_{\rm eff}^{(i)}$ at the time of each Raman pulse ($i = 1,2,3$). We then calculate the displacement vectors at the final $\pi/2$-pulse as follows:
\begin{subequations}
\begin{align}
  \label{eq:deltaChiAngle}
  \delta \bm{R}(2T) & = \frac{2\hbar T}{M} \left( \bm{k}_{\rm eff}^{(1)} - \bm{k}_{\rm eff}^{(2)} \right), \\
  \delta \bm{P}(2T) & = \hbar \left( \bm{k}_{\rm eff}^{(1)} - 2\bm{k}_{\rm eff}^{(2)} + \bm{k}_{\rm eff}^{(3)} \right), \\
  \delta \bm{\chi}(2T) & = \delta\bm{P}(2T) - \frac{M}{\Delta t} \delta\bm{R}(2T) = 
  \hbar \left( \bm{k}_{\rm eff}^{(3)} - \bm{k}_{\rm eff}^{(2)} \right),
\end{align}
\end{subequations}
where we took an expansion time of $\Delta t = 2T$ in the last expression. The contrast loss is then computed numerically using Eq.~\eqref{eq:ContrastLossCalcul}. 

\begin{figure}[!h]
  \centering
  \includegraphics[width=0.9\textwidth]{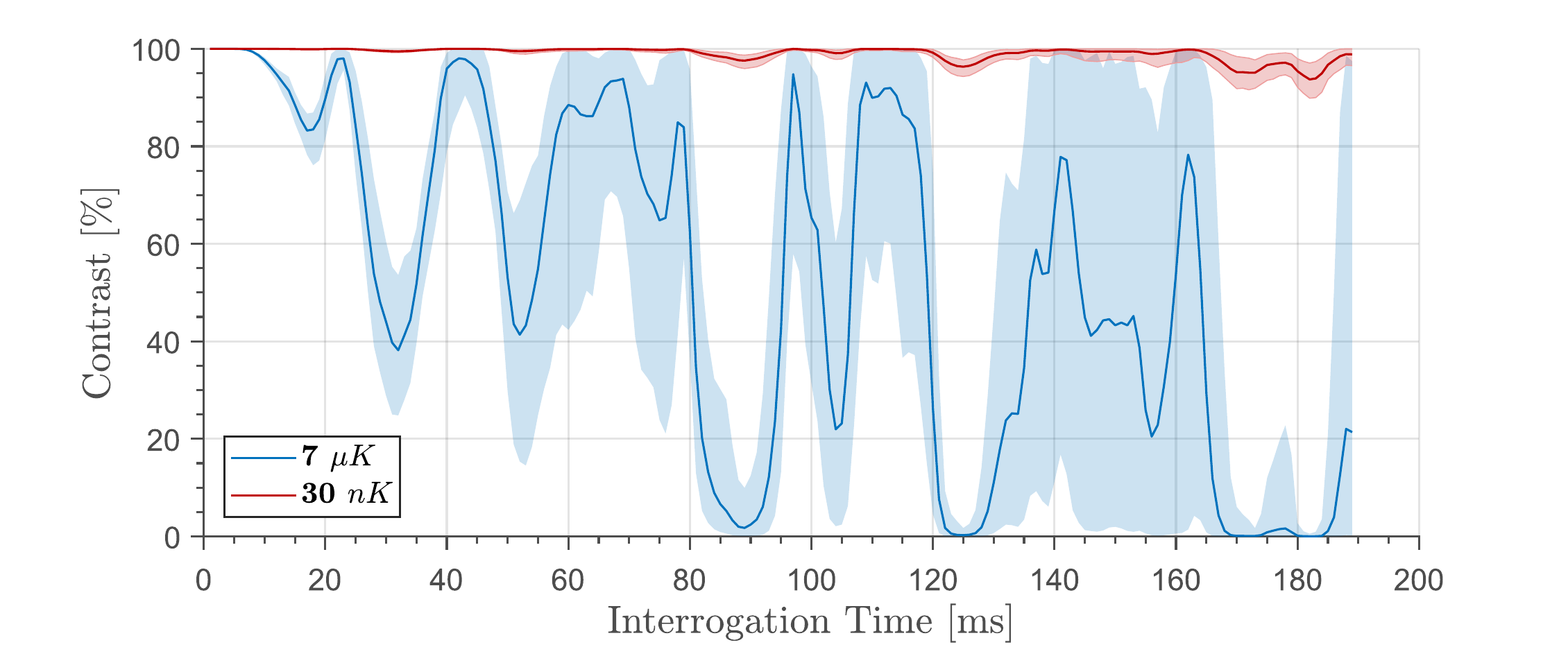}
  \caption{\textbf{Model of contrast loss during the atom interferometer.} The blue curve corresponds to a sample temperature of $\mathcal{T}= 7\mu$K. The red curve to indicates ultracold atoms at $\mathcal{T} = 30$ nK. In both cases, solid lines indicate the mean response of 100 measured platform trajectories. The shaded regions bracket the minimum and maximum variation in contrast from these trajectories. The oscillating behavior in the contrast is due to the variable rotation rate of the EE platform during its motion.}  
  \label{fig:Contrast}
\end{figure}

Figure \ref{fig:Contrast} shows our model of the contrast loss factor as a function of interrogation time $T$ at sample temperatures of $\mathcal{T} = 7$ $\mu$K and 30 nK. Our model combines the ``open interferometer'' effects of the time-varying rotation rate of the EE platform and the thermal expansion of the cloud. Our present temperature of 7 $\mu$K leads to $\sim 40\%$ contrast after only $T \simeq 30$ ms. We also predict significant fluctuations in the contrast due to the measured rotation rate noise. These effects result from mechanical vibration modes of the EE platform that are excited during its motion. The amplitude of these modes varies during the motion, and thus the contrast loss is strongly dependent on when the Raman pulses occur. This issue can be largely mitigated using ultracold atoms in an optical dipole trap \cite{Condon2019}, where temperatures of 30 nK are routinely achieved.
%

\subsection*{Bayesian estimation of the acceleration sensitivity} 

In this section, we describe a robust Bayesian method to estimate interference fringe parameters (amplitude $A$, offset $\mu_B$, and offset noise $\sigma_B$) from a set of atomic population measurements. From these estimates we accurately determine the AI signal-to-noise ratio and single-shot acceleration sensitivity. This analysis method does not rely on heuristic methods, such as non-linear least-squares fitting, nor does it require large amounts of data to estimate sample distributions from histograms. Furthermore, it converges at an optimal rate ($\sim 1/\sqrt{N}$) toward unbiased estimates of these fringe parameters.


\subsubsection*{Fringe noise model} 

The total population ratio measured at the AI output can be expressed as
\begin{equation}
  R = A \cos(\phi) + B,
\end{equation}
where $A$ is the fringe amplitude, $B$ is the fringe offset, and $\phi$ is the interferometer phase. Considering the residual vibration levels experienced on the EE platform and interrogation times of $T \gg 1$ ms, the phase $\phi$ is randomly scanned over a large interval, covering many sinusoidal periods. Therefore, after being wrapped on the $\phi \in [0, 2\pi]$ range, it can be safely assumed that random variable $\phi$ is described by a uniform probability distribution. Moreover, the fringe amplitude $A$ and offset $B$ can be considered as random variables affected by Gaussian noise distributions with mean values $\mu_A$ and $\mu_B$ and standard deviations $\sigma_A$ and $\sigma_B$, respectively.

\begin{figure}[!th]
  \centering
  \includegraphics[width=0.7\textwidth]{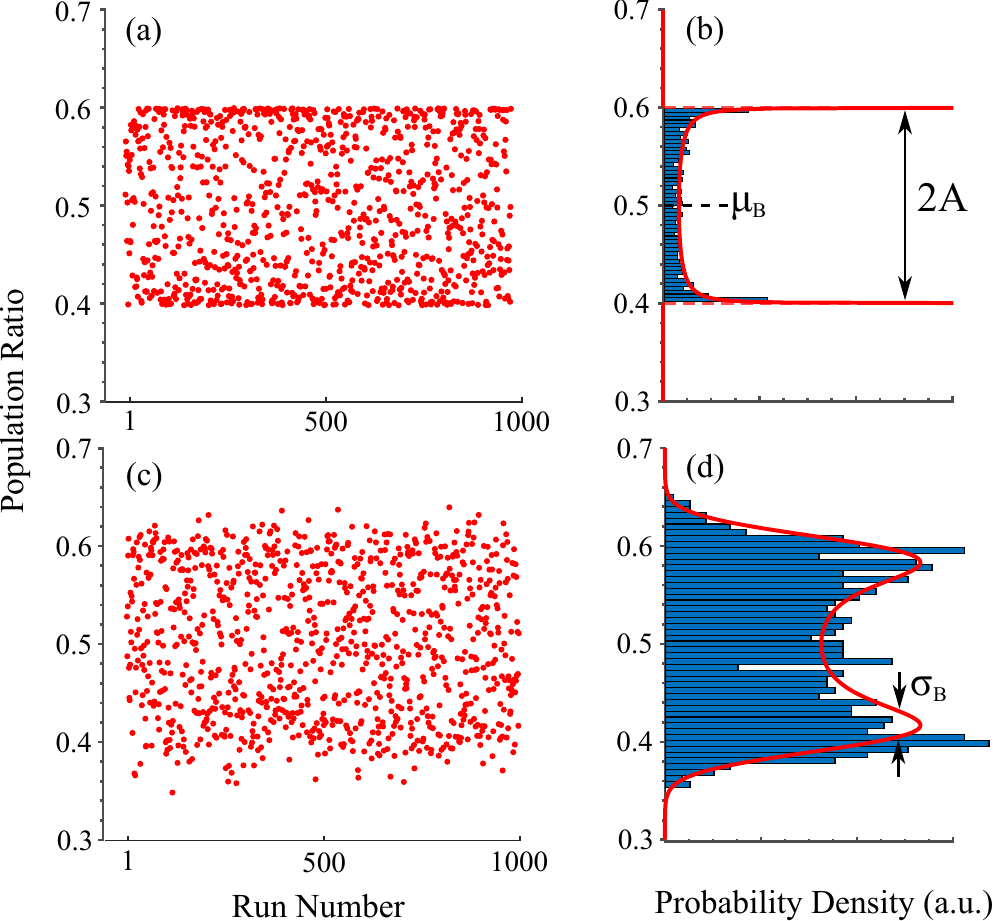}
  \caption{\textbf{Expected statistical distribution of AI measurements.} (a) Simulated output of an AI without offset noise ($A = 0.1, \mu_B = 0.5, \sigma_B = 0$). (b) The corresponding histogram of 1000 points (blue) and the PDF (red) of a pure sinusoidal function predicted by Eq.~\eqref{Sine_PDF}. (c) Simulated AI output with non-zero offset noise ($A = 0.1, \mu_B = 0.5, \sigma_B = 0.02$). (d) Corresponding histogram (blue) and PDF predicted by Eq.~\eqref{AI_Noise_Model}.}
  \label{fig:BAT_PDF}
\end{figure}

In the following simplified treatment, we consider the case where the offset noise is the dominant source of noise ($\sigma_B/\mu_B \gg \sigma_A/\mu_A$), allowing us to treat the amplitude $A$ as a constant. The more general case follows naturally from this treatment.
The measured population ratio $R$ is the sum of two \emph{independent} random variables $R = X + Y$ with $X = A \cos(\phi)$ and $Y = B$, whose probability density functions (PDFs) are respectively given by:
\begin{subequations}
\begin{align} 
  \label{Sine_PDF}
  f_X(u) & = 
  \left\{ \begin{array}{cl}
    \frac{1}{A \pi \sqrt{1-\left( u/A \right)^2}} & \quad \mbox{if } u \in ]-A, +A[, \\
    0 & \quad \mbox{otherwise,}
  \end{array} \right. \\
  \label{Offset_PDF}
  f_Y(u) & = \frac{1}{\sigma_B \sqrt{2\pi}} \exp{\left[-\frac{\left(u-\mu_B\right)^2}{2 \sigma_B^2}\right]}.
\end{align}
\end{subequations}
%
Here, $f_X(u)$ is the usual PDF of a sinusoidal function, which features a characteristic bi-modal shape with singularities at $u = \pm A$. The PDF of the sum $R$ is given by the convolution $f_R = (f_X * f_Y)$, which can be written as: 
\begin{equation}
  f_R(u) = \int f_X(z) f_Y(u - z) dz = \frac{1}{\sqrt{2} \pi^{3/2} A \sigma_B} \int_{-A}^{+A} \frac{1}{\sqrt{1-(z/A)^2}} \exp{\left(-\frac{\left(u - z - \mu_B\right)^2}{2\sigma_B^2}\right)} dz.
  \label{AI_Noise_Model}
\end{equation}
%
%
The effect of the offset noise is illustrated in Fig.~\ref{fig:BAT_PDF}. Generally, the larger the noise parameter $\sigma_B$, the more smearing of the bi-modal distribution.

\subsubsection*{Bayesian estimation algorithm} 

\begin{figure}[!th]
  \centering
  \includegraphics[width=0.8\textwidth]{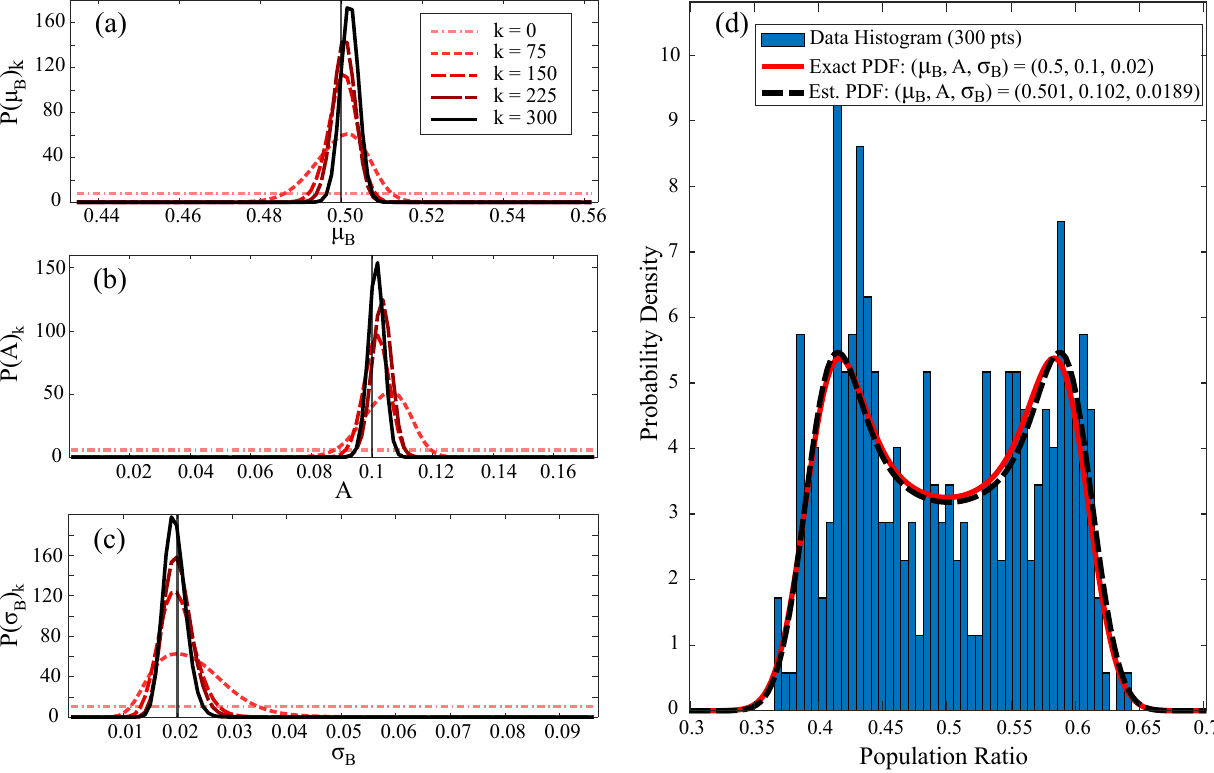}
  \caption{\textbf{Bayesian estimation of fringe parameters.} Estimation $\mu_B$, $A$, and $\sigma_B$ from a synthetic dataset of 300 points. (a), (b), and (c) shows the evolution of the probability densities associated with each parameter with measurement iteration $k$. As $k$ increases, a rapid convergence towards the real values (vertical solid lines) can be observed. (d) The final PDF derived from the estimated parameters (black dashed line) closely matches the exact PDF corresponding to the input parameters (red solid line).}
  \label{fig:Bayes_algo}
\end{figure}

Once the sensor noise model has been derived, it can be used in the frame of a Bayesian analysis in order to optimally estimate the various parameters of interest ($\mu_B$, $A$, and $\sigma_B$ in this case). The Bayesian algorithm is similar to that presented in Refs.~\cite{Stockton2007,BARRETT2015}. It consists of updating the probability distribution for the parameter state $V = \{ \mu_B, A, \sigma_B \}$ after each new measurement of $R$. An iteration $k$ of the algorithm can be decomposed into three main steps that are looped over the set of measurements $\{R\}$:
\begin{enumerate}
  \item The prior distribution for the current iteration $P(V)_k$ is set equal to the conditional distribution from the previous iteration $P(V|R_{k-1})_{k-1}$.
  \item A new measurement $R_k$ is added and the likelihood distribution $P(R_k|V)$ is computed directly from the noise model [Eq.~\eqref{AI_Noise_Model}].
  \item The updated conditional distribution after the $k^{\rm th}$ measurement is derived from Baye's rule: $P(V|R_k)_k = P(V)_k \, P(R_k|V)/N$, where $N$ is a normalizing factor.
\end{enumerate}
The application of this Bayesian analysis in a 3D parameter space is computationally demanding. We found that the execution time can be significantly reduced by using pre-computed look-up tables for the likelihood distribution $P(R_k|V)$. Starting from a uniform prior distribution $P(V)_0$ in this 3D space, the algorithm quickly converges to a much narrower PDF, from which the parameters estimates and confidence intervals can be accurately retrieved (see Fig.~\ref{fig:Bayes_algo}). These estimates and uncertainties can finally propagate to the AI signal-to-noise ratio $\text{SNR} = A/\sigma_B$ and the shot-to-shot sensitivity:
\begin{equation}
  \mathbb{S} = \frac{1}{\text{SNR} \, k_{\rm eff} \, T^2}.
\end{equation}

The Bayesian approach offers a number of advantages compared to heuristic approaches, such as least-squares fits to the sample PDF which depends on other methods like histograms or kernel density estimation. The sample PDF is itself challenging to estimate given a limited number of measurements. First, Bayesian inference makes optimal use of all available data: it doesn't suffer from the inevitable loss of information that occurs when combining multiple measurements into a histogram bin or overlapping kernels. Second, this method has no control parameters for the user to arbitrarily choose (e.g., histogram bin widths, kernel smoothing bandwidth). The Bayesian method provides an optimal, unbiased estimate of each parameter, with a statistical uncertainty that scales as $1/\sqrt{N}$. It also provides the complete likelihood distribution for each model parameter from which their confidence intervals can be calculated. Our method has been tested on both simulated and experimental data, and has proved to be extremely reliable for a broad range of fringe parameters.

 
\subsubsection*{Drift tracking} 

The approach presented above implicitly assumes that fringe parameters $\{ \mu_B, A, \sigma_B \}$ are constant quantities over time. However, in practice, instabilities in some experimental parameters, such as the detection pulse intensity, can result in slow temporal variations of the fringes parameters. 

\begin{figure}[!bh]
  \centering
  \includegraphics[width=0.7\textwidth]{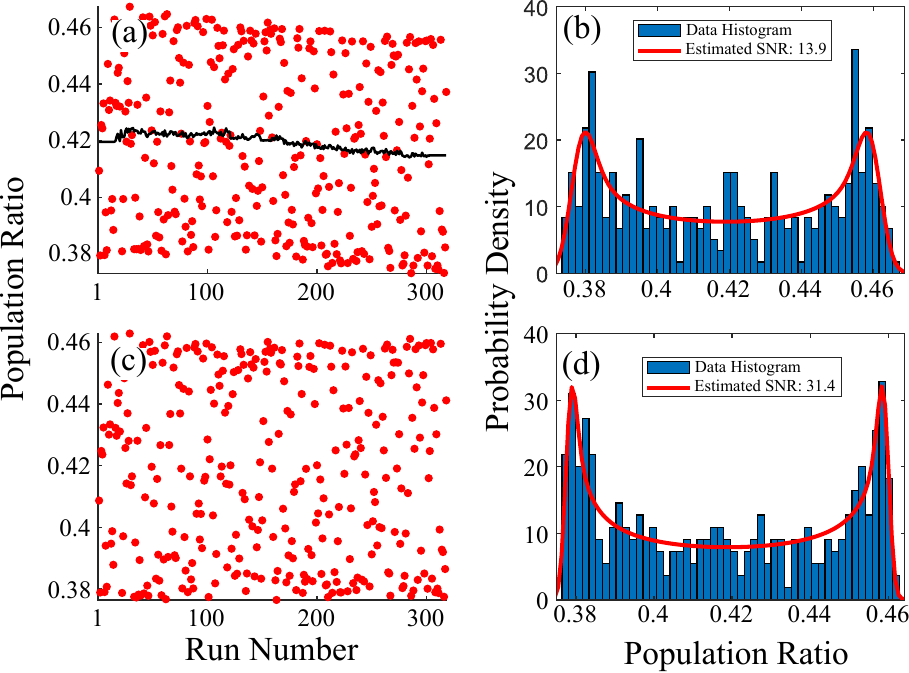}
  \caption{\textbf{Tracking offset parameter drift.} (a) Raw experimental data of 319 measurements exhibiting a slow variation of the offset parameter $\mu_B$. The drift is estimated here by using the Bayesian approach on a sliding window of 20 points (black line). (b) Histogram of raw data (blue) and estimated PDF without drift correction (SNR = 13.9). (c) Experimental data after correcting the offset drift. (d) Histogram of corrected data (blue) and estimated PDF (SNR = 31.4).}
  \label{fig:Offset_drift}
\end{figure}

Figure \ref{fig:Offset_drift} illustrates how drifts in the offset $\mu_B$ can lead to underestimates of the SNR if they are not taken into account. Since these variations are slow with respect to the cycling time of our experiment, an effective mitigation strategy consists of dynamically tracking the offset drift by applying the same Bayesian approach on a ``sliding window'', i.e., a moving subset of successive experimental data points. Thanks to its rapid convergence, the Bayesian algorithm is capable of providing a reliable estimate of the ``instantaneous'' offset value after only a few tens of measurements, making it possible to track the drift accurately, as demonstrated in Fig.~\ref{fig:Offset_drift}. This drift can then be removed from the experimental measurements before a final Bayesian analysis can be applied on the full (corrected) data set.

The efficiency of this offset tracking method was also benchmarked on simulated data and has produced better results than alternative methods based on a moving average or low-pass filtering, highlighting once again the optimality of the Bayesian approach.

\subsection*{Bayesian estimation of differential phase for equivalence principle tests} 

Experimental tests of the WEP involve precisely measuring the differential acceleration $\Delta a = a_1 - a_2$ between two test masses $M_1$ and $M_2$. These tests are characterized by the E\"{o}tvos parameter $\eta = \Delta a/a$, where $a = (a_1 + a_2)/2$ is the mean acceleration of the two masses, i.e.~the gravitational acceleration $g$. A violation of the WEP would result in a non-zero value of $\Delta a$ and, hence, of $\eta$. We consider cold atomic sources of $^{39}$K ($M_1 \simeq 38.96$ u) and $^{87}$Rb ($M_2 \simeq 86.91$ u) as test masses in a differential matter-wave accelerometer \cite{SCHLIPPERT2014, BARRETT2015, BARRETT2022}. Here, each atomic species produces an interference fringe that we model as follows:
\begin{subequations}
\label{syst1}
\begin{align}
  R_1 & = B_1 + A_1 \cos(S_1 a_1), \\
  R_2 & = B_2 + A_2 \cos(S_2 a_2),
\end{align}
\end{subequations}
where $a_j$ is the acceleration of species $j = 1,2$ relative to a common reference frame, $A_j$ and $B_j$ are the fringe amplitude and offset, respectively, and $S_j \simeq k_j T^2$ is the corresponding AI scale factor. For Raman-type AIs we use separate Raman lasers near the D2 transitions of $^{39}$K and $^{87}$Rb. The effective Raman wavevectors are $k_1 \simeq 4\pi/766.70$ rad/nm ($^{39}$K) and $k_2 \simeq 4\pi/780.24$ rad/nm ($^{87}$Rb), which leads to a scale factor ratio $\kappa = S_1/S_2 = k_1/k_2 \simeq 1.0177$. Alternatively, a Bragg-type AI employing a common laser beam at 785 nm (where the Rabi frequencies of the two species are equal \cite{Elliott2023}) yields identical scale factors ($\kappa = 1$) since $k_1 = k_2 \simeq 4\pi/785$ rad/nm. 

Equation \eqref{syst1} can be recast in a normalized form by defining a common phase $\phi_c \equiv S_2 a_2$ and differential phase $\phi_d = S_1 (a_1 - a_2) = S_1 \eta a$:
%
\begin{subequations}
\label{syst3}
\begin{align}
  n_1 & = (R_1 - B_1)/A_1 = \cos(\kappa \phi_c + \phi_d), \\
  n_2 & = (R_2 - B_2)/A_2 = \cos(\phi_c).
\end{align}
\end{subequations}
These equations describe a Lissajous curve when $n_1$ and $n_2$ are plotted parametrically. In previous work \cite{BARRETT2015}, we established a robust Bayesian method to extract the differential phase from data that follow this model. More recently \cite{BARRETT2022}, as part of a precise test of the WEP, we benchmarked the performance of this method against sinusoidal fits to each fringe (when the interferometer phase is known). As input, the algorithm requires estimates of the common phase range $[\phi_c^{\rm min}, \phi_c^{\rm max}]$, and the relative noise parameters for each atomic species: $\sigma_{A,j}/A_j$ and $\sigma_{B,j}/A_j$. In our case, the motion of the inertial reference frame determines the common phase range, while the noise parameters can be estimated from a sample of population ratios $R_1$ and $R_2$ using the novel Bayesian method described in the previous section.

The simulations in Figure \ref{fig:BayesianSimulator} shows illustrate the sensitivity of the differential phase estimate at current noise levels on the EE platform. We chose $\phi_d^{\rm true} = 1.5$ rad for illustrative purposes. Using the measured vibration spectrum of the EE platform, we estimate a common phase range of $\phi_c \in [-15,15]$ rad for a total interrogation time $2T = 180$ ms. This produces many crossings in the Lissajous curves for the Raman case. We included a relative offset noise of $\sigma_{B,1}/A_1 = \sigma_{B,2}/A_2 = 0.06$ and a differential phase noise of $\sigma_{\phi_d} = 80$ mrad (determined from previous measurements \cite{BARRETT2022}). Noise in the fringe amplitudes was ignored in these simulations ($\sigma_{A,1} = \sigma_{A,2} = 0$). For $N > 10$ measurements, this Bayesian method yields an optimal scaling of $\sim 1/\sqrt{N}$ for the statistical uncertainty on the differential phase $\Delta \phi_d^{\rm stat}$. Importantly, the Bayesian method is also unbiased---the mean systematic error (black dots on Figure \ref{fig:BayesianSimulator} (c) and (f)) $\Delta \phi_d^{\rm sys} = \langle|\phi_d^{\rm Bayes} - \phi_d^{\rm true}|\rangle$ is always less than the statistical one.

\section*{Acknowledgments} 

This work is supported by the French national agency CNES (Centre National d’Etudes Spatiales), the European Space Agency (ESA), and co-funded by the European Union (CARIOQA-PMP project) and PEPR (Programmes et Equipements Prioritaires de Recherche) France Relance 2030 QAFCA grant no.~ANR-22-PETQ-0005 QAFCA. For financial support, C. Pelluet and C. Metayer thank CNES, CNRS, and ESA. B. Barrett thanks NSERC (Natural Science and Engineering Research Council of Canada), the Canadian Foundation of Innovation (CFI), and the New Brunswick Innovation Foundation (NBIF) for financial support. 

\section*{Author contributions} 

P.B. and B.Battelier conceived the project. M.R. and B.Barrett built the apparatus. C.P., M.R., V.J., R.A., and C.M. performed experiments, C.P., B.Barrett and R.A. carried out the data analysis. B.Battelier supervised the experiments and data analysis. B.Battelier coordinated and administrated the project. B.Battelier, C.P., R.A., P.B., and B.Barrett wrote the manuscript. All authors discussed and reviewed the manuscript.

\section*{Data availability} 

The data that support the findings of this study are available from the corresponding author upon request.

\section*{Competing interests} 

The authors declare no competing interests.



\end{document}